\documentclass[runningheads]{svmult}

\usepackage{makeidx}   
\usepackage{graphicx}  
\usepackage{subeqnar}  
\usepackage{multicol}  
\usepackage{physprbb}  
\makeindex             

\begin{document}
\title*{Convective Patterns in Binary Fluid Mixtures with Positive Separation
Ratios}
\toctitle{Convective Patterns in Binary Fluid Mixtures with Positive Separation
Ratios}
%
%
\titlerunning{Convective Patterns in Binary Fluid Mixtures}
%
\author{Bj\"orn Huke
\and Manfred L\"ucke}
\authorrunning{Bj\"orn Huke and Manfred L\"ucke}
%
%
\institute{Institut f\"ur Theoretische Physik, Universit\"at des Saarlandes,
           Postfach 151150, 
	   D-66041 Saarbr\"ucken, Germany}

\maketitle              

\begin{abstract}

We summarize our findings about laterally periodic convection structures in
binary mixtures in the Rayleigh--B\'enard system for positive Soret effect. 
Stationary roll, square, and crossroll solutions and their stability are
determined with a multimode Galerkin expansion. The oscillatory competition of
squares and rolls in the form of crossroll oscillations is reviewed. They
undergo a subharmonic bifurcation cascade where the oscillation period grows in
integer steps as a consequence of an entrainment process.
\end{abstract}

\section{Introduction}

The Rayleigh--B\'enard system is a prominent example for studying 
\index{Rayleigh--B\'enard system}
pattern formation in hydrodynamic systems driven away from equilibrium. 
\index{pattern formation}
In this system
a fluid layer is confined between two extended plates perpendicular to the
direction of gravity. A temperature difference between lower and upper plate is
applied. Below a critical temperature difference a quiescent conductive state 
is established wherein the temperature varies linearly between the plates. But 
for larger differences buoyancy forces are strong enough to destabilize this
conductive state and to start convection.
\index{convection}

The pure fluid convection in the Rayleigh--B\'enard system at moderate heating
\index{Rayleigh--B\'enard system}
\index{convection}
rates is experimentally and theoretically well investigated \cite{CH93}.
An extension of the problem that leads to more complex convection behavior
is achieved by using binary mixtures such as ethanol-water instead. In binary 
mixtures the buoyancy forces are also influenced by concentration variations.
The structural dynamics of the concentration 
distribution in mixtures results from an interplay between three competing 
mechanisms: nonlinear 
advection and mixing, weak solutal diffusion, and the Soret effect. The latter
generates and sustains concentration gradients in linear response to  
local temperature gradients. Without Soret coupling measured by the separation
ratio any concentration fluctuation diffuses away.

Convection in binary mixtures shows a rich spectrum of 
\index{convection}
pattern formation behavior \cite{CH93,PL84,LBBFHJ98}. But the knowledge about 
\index{pattern formation} 
these structures and in particular about their stability is more
\index{stability}
limited than for pure fluids. This is especially true for the case of a 
positive Soret effect where the fact that the novel structures are
three-dimensional makes their numerical investigation more difficult.  

A positive separation ratio implies that temperature differences drive 
the lighter component of the mixture into the direction 
of higher temperature. That means that the Soret effect enhances the buoyancy
forces that result from the temperature dependence of the density. Convection 
\index{convection}
is therefore established at temperature differences smaller than those needed 
for pure fluids. The interval of temperature differences that allows convection in a
binary mixture but not yet in a pure fluid is called Soret region. 
Therein convection is driven only because of the presence of the solutal
contribution to the buoyancy force. Above this range, i.~e. in the
Rayleigh region, the thermal part of the buoyancy is most important and the
Soret effect is less effective for destabilizing the conductive state.
\index{stability}

In the Soret region square-like convection patterns are often
\index{convection}
\index{squares}
found as the primary stable form of convection whereas in
\index{convection}
\index{stability}
pure fluids roll patterns are stable at onset. Such rolls can also be
\index{rolls}
found in binary mixtures, especially at higher heating rates where convection is
strong enough to reduce the concentration variations sufficiently by advective 
mixing. In between there exist two different kinds of crossroll 
\index{crossrolls}
structures: one is stationary and the other one is oscillatory. They are stable
where neither rolls nor squares are \cite{GPC85,MS91,DAC95,JHL98}.
\index{squares}
\index{rolls}
\index{stability}
In this article we summarize our findings about laterally periodic 
convection structures for positive separation ratios.
\index{convection}
In Sect.~\ref{technique} we present the basic equations for binary mixture
convection in the Rayleigh--B\'enard system and the Galerkin technique we used
\index{Galerkin method}
\index{Rayleigh--B\'enard system}
to investigate the different convection structures numerically. In 
Sects.~\ref{rolls} -- \ref{crossrolls} we discuss the properties of rolls,
\index{rolls}
squares, and crossrolls and their range of stability. We conclude in
Sect.~\ref{conclude}.
\index{stability}
\index{crossrolls}  
\index{squares}

\section{Numerical Methods}
\label{technique}

\subsection{System and Basic Equations}
\label{sec21}

We consider a horizontal layer of a binary fluid mixture of thickness $d$ in
a homogeneous gravitational field, $\vec{g} = - g \, \vec{e}_z$.
A vertical temperature gradient is imposed by fixing the temperature
\begin{equation}
T = T_0 \pm \frac{\Delta T}{2} 
\,\,\, \mbox{at} \,\,\, z = \mp \frac{d}{2} \, ,
\end{equation}
e.g., via highly conducting plates in experiments. Here we consider the 
plates to be infinitely extended, rigid, and impermeable. $T_0$ is the mean
temperature of the fluid layer.

In the conductive state a linear temperature profile 
\begin{equation}
T_\mathrm{cond}(z) = T_0 - \frac{\Delta T}{d} z 
\end{equation}
is established. If there is a Soret effect, the temperature gradients generate
also a concentration gradient:
\begin{equation}
C_\mathrm{cond}(z) = C_0 + S_\mathrm{T} C_0 (1 - C_0) \frac{\Delta T}{d} z \; .
\end{equation}
Here $C$ means the concentration of the lighter component with $S_\mathrm{T}$ 
being its Soret coefficient. We will consider the case of negative 
$S_\mathrm{T}$.
In this case the lighter component of the mixture is driven into the direction 
of higher temperature. Concentration at the bottom is 
larger in the conductive state for positive $\Delta T$ 
therefore enhancing the density gradient and further destabilizing the 
\index{stability}
fluid layer.

Convection is described in terms of the fields of $T$, $C$, velocity 
\index{convection}
$\vec{u} = (u,v,w)$, total mass 
density $\varrho$, and pressure $P$. In the balance equations connecting 
these fields we scale lengths and positions by $d$, time by the 
vertical thermal diffusion time $d^2 / D_\mathrm{th}$, temperature by 
$\nu D_\mathrm{th} / \alpha g d^3$, concentration by 
$\nu D_\mathrm{th} / \beta g d^3$, 
and pressure by $\varrho_0 D_\mathrm{th}^2/d^2$. Here $\varrho_0$ is the 
mean density, $D_\mathrm{th}$ the thermal diffusivity, $\nu$ the kinematic 
viscosity, and $\alpha = -(1/\rho) \partial \rho/ \partial T$ and 
$\beta = -(1/\rho) \partial \rho/ \partial C$ are thermal and solutal expansion
coefficients, respectively.
Using the Oberbeck--Boussinesq approximation 
the balance equations read \cite{PL84,HLL92}
\begin{subeqnarray}
\nabla \cdot \vec{u} & = & 0 \label{ground21}\\
(\partial_t + \vec{u} \cdot \nabla \,) \, \vec{u} & = & 
- \nabla \,p  +  
Pr\,\left[\,\left(\theta + c\,\right)\, \vec{e}_z + 
\nabla^2\, \vec{u} \,\right] \label{ground22}\\
(\partial_t + { \vec{u} \cdot \nabla \,})\,\theta & = & 
Ra\,w  + \nabla^2\,\theta\label{ground23}\\
(\partial_t + \vec{u} \cdot \nabla\,)\,c & = & Ra\,\psi\,w  +  
Le\,\left(\,\nabla^2\,c - \psi\,\nabla^2\,\theta\,\right)\label{ground24} \; .
\end{subeqnarray}
Here $\theta, c$, and $p$ are the reduced deviations of temperature, 
concentration, and pressure, respectively, from the conductive profiles.                              

The Lewis number $Le$ is the ratio of the
concentration diffusivity $D$ to the thermal diffusivity $D_\mathrm{th}$, 
therefore
measuring the velocity of concentration diffusion. The Prandtl number 
$Pr$ is the ratio of the momentum diffusivity $\nu$ and $D_\mathrm{th}$:
\begin{equation}
Le = \frac{D}{D_\mathrm{th}}\,\,;\,\, Pr = \frac{\nu}{D_\mathrm{th}} 
\label{GdefLsig} \,.
\end{equation}
The Rayleigh number $Ra$ measures the thermal driving and the separation ratio
$\psi$ measures the strength of the Soret coupling between temperature and 
concentration fields
\begin{equation}
Ra =\frac{ \alpha g d^3 \Delta T}{\nu D_\mathrm{th}}\,\,;
\,\psi = -\frac{\beta}{\alpha} S_\mathrm{T} C_0 ( 1 - C_0) \, .
\label{GdefR} 
\end{equation}
We consider here $S_\mathrm{T}<0$, i.~e.~$\Psi >0$ for mixtures with 
$\alpha,\beta>0$. 

The first equation (\ref{ground21}) expresses the fact that the fluid is 
considered to be incompressible. (\ref{ground22}) -- (\ref{ground24}) are the
equations of motion for $\vec{u}$, $\theta$ and $c$. The left hand sides of 
these equations are the substantial time derivatives of the respective fields.
The driving forces entering into 
the momentum balance equation (\ref{ground22}) are pressure 
gradients and the buoyancy caused by the temperature and concentration 
dependence of the density. The remaining term on the right hand side is the
momentum diffusion term. 

(\ref{ground23}) and (\ref{ground24}) contain also diffusion terms on the right
hand side. The off-diagonal term 
$-Le\,\psi\,\nabla^2\,\theta\,$ and the term $Ra\,\psi\,w$
in the concentration balance equation (\ref{ground24})
describe the action of the Soret effect, i.e., the generation of concentration 
currents and concentration gradients by temperature 
variations.

The Dufour effect, i.e., the driving 
of temperature currents by concentration variations is of interest only in 
gas mixtures \cite{HLL92}. But even there 
it is often small \cite{LA97}. 

\subsection{Galerkin Method}
\label{sec22}
\index{Galerkin method}

We are concerned with three-dimensional convection patterns periodic in the 
\index{convection}
$x$- and $y$-direction. To describe such patterns with wavenumbers $k_x$ 
and $k_y$ each field $X$ is expanded as
\begin{equation}
X(x,y,z;t) = \sum_{lmn} X_{lmn}(t) \E^{\I l k_x x} \E^{ \I m k_y y } f_n(z) 
\; .
\label{expansion}
\end{equation}
Here $l$ and $m$ are integers and the $f_n$ form a complete system of 
functions that fits the specific boundary condition for the field $X$ at the
plates. 
To find suitable sets of functions $f_n$ we introduce some new fields.
First, two scalar fields $\Phi$ and $\Psi$ are defined via
\begin{equation}
\vec{u} = \nabla \times \nabla \times \Phi \vec{e}_z
+ \nabla \times \Psi \vec{e}_z \; . \label{22phipsidef}
\end{equation}
The structures we want to discuss do not show a horizontal mean flow for 
mirror symmetry reasons. Then, (\ref{22phipsidef})
is the most general expression that fulfills the incompressibility
condition (\ref{ground21}) \cite{CB89}.

Second, instead of $c$ we use the field 
\begin{equation}
\zeta = c - \psi \theta
\end{equation}
that allows in a more convenient way to guarantee the impermeability of 
the horizontal boundaries: The diffusive part of the concentration current, 
driven by concentration gradients as well as by temperature
gradients is given by $-Le \, \nabla(c-\psi\theta)$. 
At the impermeable plates the vertical component of this current vanishes 
which requires
\begin{equation}
0 = \partial_z \left(c - \psi \theta \right) = 
    \partial_z \zeta  \hspace{1cm} 
\mbox{at } z = \pm 1/2 \; .
\end{equation}
The advective concentration current vanishes at the plates because there 
${\bf u} = 0$.
The balance equation for $\zeta$ is obtained by combining (\ref{ground23}) and
(\ref{ground24}).

The boundary conditions for the fields $\Phi, \Psi, \theta$, and 
$\zeta$ read
\begin{equation}
\Phi = \partial_z \Phi = \Psi = \theta = \partial_z \zeta = 0 
\hspace{2cm} \mbox{at } z = \pm 1/2 \; .
\end{equation}
To expand the fields 
$\Psi$, $\theta, \zeta$, and $\Phi$ vertically we used different 
orthonormal sets $f_n(z)$ as follows
\begin{subeqnarray}
\Psi \mbox{\, and \,} \theta \,:\,f_n(z) &=& \, 
\left\{ \begin{array}{ll}
\sqrt{2} \cos(n \pi z)  & \mbox{\qquad $n$ odd} \\ 
\sqrt{2} \sin(n \pi z)  & \mbox{\qquad $n$ even} 
\end{array} \right.
\label{csdefinition} \\
\zeta \,:\, f_n(z)  &=& \left\{ \begin{array}{ll}
1 & \mbox{\qquad  $n = 0$}\\
\sqrt{2} \, \sin(n \pi z) & \mbox{\qquad $n$ odd} \\
\sqrt{2} \, \cos(n \pi z) & \mbox{\qquad $n \neq 0$ even}
\end{array}
\right.
\label{scdefinition} \\
\Phi \,:\, f_n(z) &=& \left\{ \begin{array}{ll} 
C_{\frac{n+1}{2}}(z) & \mbox{\qquad $n$ odd} \\ 
S_{\frac{n}{2}}(z) & \mbox{\qquad $n$ even} 
\end{array} \right. \; .
\label{CSdefinition}
\end{subeqnarray}
Here $C_n$ and $S_n$ are Chandrasekhar--Reid functions \cite{C81}.

The balance equations for the new fields are
\begin{subeqnarray}
\partial_t \Delta_2 \Psi & = & 
Pr \nabla^2 \Delta_2 \Psi + 
\left\{\nabla \! \times \left[ \left(\vec{u} \cdot \nabla \right) 
\vec{u} \right] \right\}_z  \label{ground71}\\
\partial_t \nabla^2 \Delta_2 \Phi & = & 
Pr \left\{ \nabla^4 \Delta_2 \Phi -  
\Delta_2 \left[ \left( 1 + \psi \right) \theta + \zeta \right] \right\} 
\nonumber \\
&&
- \left\{\nabla \! \times \nabla \! \times
\left[ \left( \vec{u} \cdot \nabla \right) \vec{u} \right] \right\}_z 
\label{ground72} \\
(\partial_t +\vec{u} \cdot \nabla )\,\theta & = & 
-Ra\, \Delta_2 \Phi  + \nabla^2\,\theta \label{ground73} \\
(\partial_t + \vec{u} \cdot \nabla )\,\zeta & = & 
Le \nabla^2 \zeta - \psi \nabla^2 \theta \label{ground74}
\; .
\end{subeqnarray}
Here $\Delta_2 = \partial_x^2 + \partial_y^2$.

By inserting the ansatz (\ref{expansion}) for each field into the balance 
equations and projecting them 
onto the basic functions one gets a nonlinear algebraic system of equations of 
the form
\begin{equation}
A_{\kappa \mu} \partial_t X_\mu = B_{\kappa \mu} X_\mu + 
C_{\kappa \mu \nu} X_\mu X_\nu \; . \label{basicslinsyseq}
\end{equation} 
For simplicity amplitudes are labelled here by a single Greek 
index and the summation convention is implied in (\ref{basicslinsyseq}) with
$A_{\kappa \mu}, B_{\kappa \mu}$, and $C_{\kappa \mu \nu}$ being constant
 coefficients.
 
The number of modes has to be truncated to get a 
finite number of equations as discussed later on. For stationary 
convection structures the 
\index{convection}
left hand side of (\ref{basicslinsyseq}) vanishes and the solution can be 
found using a multidimensional Newton method.

\subsection{Stability Analysis}
\index{stability}
\label{sec24}

To make a full stability analysis one has to check the stability of the 
\index{stability}
patterns against perturbations with arbitrary wavevector 
$d \vec{e}_x + b \vec{e}_y $. To do so one has
to introduce a Floquet term writing out the perturbation as
\begin{equation}
\delta X(x,y,z;t) = \E^{\I d x + \I b y} \E^{s t}
\sum_{lmn} \delta X_{lmn}  \E^{\I l k_x x + 
\I m k_y y} f_n(z) \; .
\end{equation}
Such a perturbation is added to the known solution the stability of which is to 
\index{stability}
be tested and inserted into the 
balance equations. After linearizing and projecting one gets a linear eigenvalue
problem of the form
\begin{equation}
s A_{\kappa \mu} \delta X_\mu = B_{\kappa \mu} \delta X_\mu \; . 
\end{equation}
with constant coefficients $A_{\kappa \mu}$ and $B_{\kappa \mu}$. 
The aforementioned solution, i.e., the convective structure described by it 
is stable if 
every eigenvalue $s$  has a negative real part for every $d$ and $b$.
\index{stability}

The symmetry of the convective pattern discussed above can under some 
circumstances be 
used to get separated classes of possible eigenvectors representing the 
perturbations. That means the eigenvalue problem can be reduced 
to finding the eigenvalues of two matrices of about half of the size. Because 
evaluating the eigenvalues of a matrix is a $O(N^3)$-process this always implies 
a reduction of the computation time.

\subsection{Parameter Space}

In pure fluids convection in the Rayleigh--B\'enard system starts at a critical
\index{Rayleigh--B\'enard system}
\index{convection}
Rayleigh number $Ra_\mathrm{c}^0 =Ra_\mathrm{c}(\psi=0)=1707.762$ and a 
wavenumber $k_\mathrm{c}^0 =k_\mathrm{c}(\psi=0) = 3.117$ \cite{C81}. In binary 
mixtures 
with positive separation ratios that we are dealing with, the critical Rayleigh
number is smaller, $Ra_\mathrm{c}(\psi>0) < Ra_\mathrm{c}^0$, since the 
solutal contribution
to the quiescent state's buoyancy force enhances the latter. Thus a smaller
thermal driving, i.e., a smaller  Rayleigh number suffices to reach the
critical buoyancy force size for onset of convection.
\index{convection}
The critical wavenumber is also somewhat lower: 
$k_\mathrm{c}(\psi>0) < k_\mathrm{c}^0$ 
\cite{KM88}. 
When presenting our results we shall use the reduced
Rayleigh number
\begin{equation}
r = \frac{Ra}{Ra_\mathrm{c}^0} \; .
\end{equation}

The convection above onset depends on the three parameters $\psi$, $Le$, and
\index{convection}
$Pr$. These parameters depend on the components of the mixture and the mean
values of temperature and concentration. For alcohol-water for example, 
$\psi$ may assume values between $-0.6$ and $0.3$. Here the Lewis number 
$Le = 0.01$ or smaller whereas $Le \approx 1$ is typical for gas mixtures. A
typical Prandtl number for liquid mixtures is $Pr=10$, in gas mixtures 
$Pr=0.1$...$1$.    

\section{Rolls}
\label{rolls}
\index{rolls}

In pure fluids, convection in the form of parallel rolls is a stable form of 
\index{convection}
\index{stability}
\index{rolls}
convection directly above onset for all values of $Pr$ \cite{B78}, 
although they compete with spiral defect chaos at low Prandtl numbers where 
the latter structure has a larger basin of attraction \cite{LA96}. In 
binary mixtures, they exist also as stable patterns above $Ra_\mathrm{c}$ for 
\index{stability}
all positive $\psi$ if $Le$ is large enough to equilibrate the concentration 
sufficiently. The required value of $Le$ depends mainly on the strength of the 
Soret effect but also on the Prandtl number. The rolls lose stability against 
squares at onset if $Le$ is too small, especially for strong Soret coupling 
\cite{ClK91}. 
\index{stability}
\index{rolls}
\index{squares}

\subsection{Technical Remarks}

Calculating the fixed point solutions and analyzing their stability requires
\index{stability}
less numerical effort than for 3D structures. Only $X_{l0n}$-modes have to be 
taken into account and $\Psi \equiv 0$. Mirror symmetry in the lateral direction
allows to set $X_{l0n} = X_{-l0n}$ so that the lateral 
functions $\E^{\pm \I k l x}$ can be replaced by 
$\cos (k l x)$. In addition the roll pattern is antisymmetric under reflection 
\index{rolls}
at the plane $z=0$ combined with a translation by half a wavelength in 
$x$-direction. This mirror glide symmetry enforces 
half of the amplitudes to be zero, e.~g.\ all amplitudes 
$\Phi_{l0n}$ where $l+n$ is an odd number.

To perform the stability analysis of rolls one determines the growth behavior
\index{stability}
\index{rolls}
of perturbations of the form
\begin{equation}
\delta X(x,y,z;t) =  \E^{\I d x + \I b y} \E^{s t} \sum_{ln} \delta X_{l0n}  
\E^{\I l k x} f_n(z) \; .
\label{perturbations}
\end{equation}

Because of the periodicity of the patterns in $x$ and its mirror symmetry 
it suffices to consider $d \in \left[ 0, k/2 \right]$. In 
$y$-direction, however, all perturbation wavenumbers, say, $b \ge 0$ 
have to be investigated. For a discussion of the effect of mean flow
perturbations, e.~g., $\delta \Phi_{00n}$ we refer to \cite{HLBJ00}.

The linear system of 
equations (\ref{perturbations}) always separates into two subsystems of
perturbations $\delta X_{l0n}$ that belong to modes with amplitudes $X_{l0n}$ 
that are antisymmetric or symmetric under the mirror glide operation 
$(x, z) \rightarrow (x + \lambda/2, -z)$.
E.~g.\ all 
perturbations with amplitudes $\delta \Phi_{l0n}$ with even $l+n$ belong to one
set, while the perturbations with odd $l+n$ belong to the other.

In special cases the system of equations can be separated even further. For
$d=0$ the perturbations can be divided into those that are symmetric and those
that are antisymmetric under the operation $x \rightarrow -x$. Furthermore, 
if $b=0$ then the perturbations contain either no or only
$\delta \Psi$-amplitudes. This is also of practical interest, since some 
instabilities are most critical in these special cases.
\index{stability} \index{instability}

The sets of modes taken into account in the Galerkin procedure were chosen as 
\index{Galerkin method}
follows: We defined a maximal mode index $N$ and neglected all modes 
$X_{l0n}$ or $\delta X_{l0n}$ with $|l|+n > N_1$ for $\Phi$- and  
$\Psi$-fields and $|l|+n > N_2$ for $\theta$- and  $\zeta$-fields. Here
$N_2 = 2 N_1$.

\subsection{Numerical Results}

A quantitative description of the fixed point solutions and the stability 
\index{stability}
behavior in binary mixtures requires in some cases sets of modes that are 
much larger than those needed for pure fluids. This is because the strong
anharmonic narrow boundary layer behavior of the $\zeta$-field for small $Le$ 
and large $r$.
For the most anharmonic roll structures at $r \approx 1.5$, $Le < 0.01$, and 
\index{rolls}
$\psi = 0.15$ that we have investigated expansions up to $N_2=40$ were needed. 
Although the velocity and temperature field are much smoother, a consistent 
description of the latter requires then also high $\theta$-modes
as discussed in \cite{HL98,H96}.
This is much more than for pure fluids, where truncations with $N \leq 8$ are 
sufficient to describe the stability behavior quantitatively even at large $r$.
\index{stability}

\subsection*{Roll Solutions}
\index{rolls}

\begin{figure}[t]
\begin{center}
\includegraphics[width=.7\textwidth,angle=270]{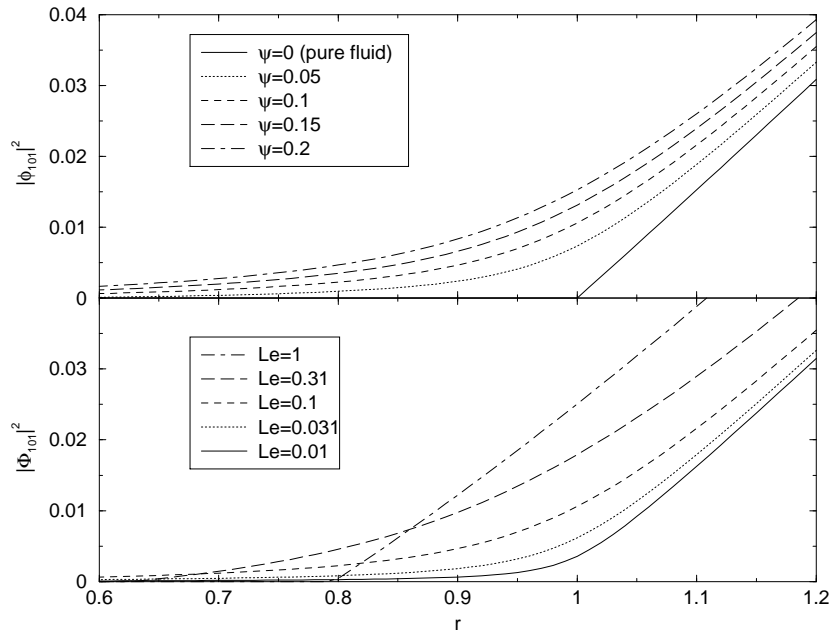}
\end{center}
\caption[]
{Mode intensity $|\Phi_{101}|^2$ of roll solutions as a function of $r$ for 
\index{rolls}
$k=k_\mathrm{c}^0$ and
$Pr=10$. Above: for $Le=0.1$ and different $\psi$. Below: for $\psi=0.1$ 
and different $Le$}
\label{amplis}
\end{figure}
We will discuss the fixed point solutions by focusing on the mode intensity
$|\Phi_{101}|^2$ at $k = k^0_\mathrm{c}$ as order parameter (see Fig.~\ref{amplis}). 
The relation to the vertical velocity $w$ is 
$k^2 \Phi_{101} = w_{101}$. It is useful to start with 
considering solutions for small $Le$. In this case, two 
different convection regimes can be distinguished. For $r < 1$, in the Soret 
\index{convection}
regime, the applied temperature gradients cause convection only indirectly, by 
generating concentration and therefore density gradients via the Soret effect. 
The convection amplitudes are very small here. This changes in the Rayleigh 
regime, where the temperature gradients cause the necessary density gradients 
to develop convection also directly. $|\Phi_{101}|^2(r)$ strongly curves upward
and $|\Phi_{101}|^2$ assumes values comparable to those of pure fluid
convection, since concentration is more and more equilibrated. 
\index{convection}

The transition between Soret and Rayleigh region is especially sharp if $\psi$
is also small. It is also present for larger $Le$, but more smooth. It vanishes 
for $Le \approx 1$. The knowledge of these features will be helpful for a 
qualitative understanding of the stability behavior.
\index{stability}

\subsection*{Stability Properties of Rolls}
\index{stability}
\index{rolls}

The stability boundaries of roll patterns in pure fluids are known since the
\index{stability}
\index{rolls}
pioneering work of Busse and his coworkers \cite{B78}.
At small Rayleigh numbers there exist five different instability
\index{stability} \index{instability}
mechanisms giving rise to five different stability boundaries 
that limit the region of stable rolls in the $(Ra,k,Pr)$-parameter space. 
\index{rolls}
At small Prandtl numbers the Eckhaus (EC), the skewed varicose 
(SV), and the oscillatory mechanism (OS) are the important instabilities 
\cite{CB90}. 
\index{stability} \index{instability}
At higher Prandtl numbers the zigzag (ZZ) and the crossroll (CR) mechanisms 
\index{crossrolls}
dominate \cite{B67}. Properties and symmetries of these perturbations 
are discussed in \cite{BBC85}. These five instabilities of roll patterns 
\index{rolls}
are the same as those that can be found in binary mixtures. A more detailed
discussion of the symmetries and other features of these mechanisms can be found
in \cite{HLBJ00}.
\index{stability} \index{instability}

\begin{figure}[t]
\begin{center}
\includegraphics[width=.7\textwidth]{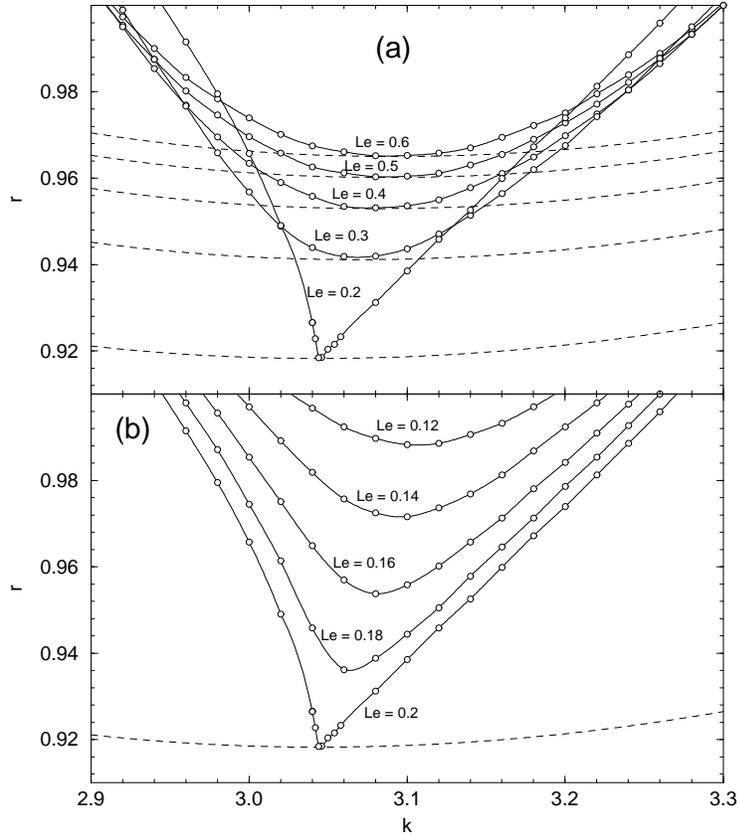}
\end{center}
\caption[]
{CR instability boundaries (solid lines) of rolls in the Soret region for 
\index{stability} \index{instability}
\index{rolls}
$Pr = 10$, $\psi = 0.01$ and several values of $Le$. Rolls are stable 
\index{rolls}
against CR-perturbations above the solid lines. (a): For $Le \geq 0.2$ the 
CR boundaries touch the neutral curve (dashed lines) in the critical point. 
(b): For $Le < 0.2$ the neutral curve goes further down (not shown) and is 
disconnected from the CR boundaries. Then rolls are not stable at the critical 
\index{rolls}
point anymore}
\label{CRevolution}
\end{figure}

The central qualitative difference between roll structures in pure fluids and
\index{rolls}
binary mixtures concerning their stability is certainly their loss of stability 
\index{stability}
in mixtures at onset in a wide range of parameter space. However, roll 
\index{rolls}
structures do in general still exist in mixtures
as stable structures for these parameters, not at onset, but at larger
\index{stability}
$r$. Figure~\ref{CRevolution} shows the CR instability boundaries of rolls
\index{rolls}
\index{instability}
in a parameter interval where an exchange of stability between rolls and 
\index{rolls}
squares at onset is predicted in \cite{ClK91}. One 
\index{squares}
sees that the curvature of the CR boundary at the critical point
diverges -- a feature that follows also from cubic amplitude
equations, see \cite{HLBJ00} -- 
when this exchange occurs. For the parameters $Pr =10, \psi=0.01$ of 
Fig.~\ref{CRevolution} the exchange occurs at $Le=0.2$. Decreasing
$Le$ below this value the neutral stability curve 
\index{stability}
(dashed line in Fig.~\ref{CRevolution}) drops further down in $r$ 
(not shown in Fig.~\ref{CRevolution}b) while the CR instability boundary
disconnects from the neutral stability curve and moves up in $r$. Above the 
CR boundary rolls are still stable against CR perturbations.
\index{rolls}
\index{stability} \index{instability}

The rolls could still be unstable there against other perturbations but we 
\index{rolls}
\index{stability}
found the minimum of the CR boundary always to be the minimal Rayleigh number
above which stable rolls exist. Only at $\psi = 0.15$, $Pr < 0.2$ and 
\index{rolls}
$Le < 0.015$ we found regions where rolls seem to be unstable everywhere in 
\index{rolls}
the $(k,r)$-plane. The two branches of the CR boundary meet again at 
higher $r$ near this region and limit an oval region of CR-stable rolls from 
\index{rolls}
below and also {\em above}. The oval region gets smaller by reducing $Le$ or
$Pr$ until the region of stable rolls vanishes.
\index{rolls}
\index{stability}
The experimental observation of this behavior might be
difficult because it occurs in a region of the parameter space that is not
accessible by ordinary fluid mixtures.

\begin{figure}[t]
\begin{center}
\includegraphics[width=.7\textwidth,angle=270]{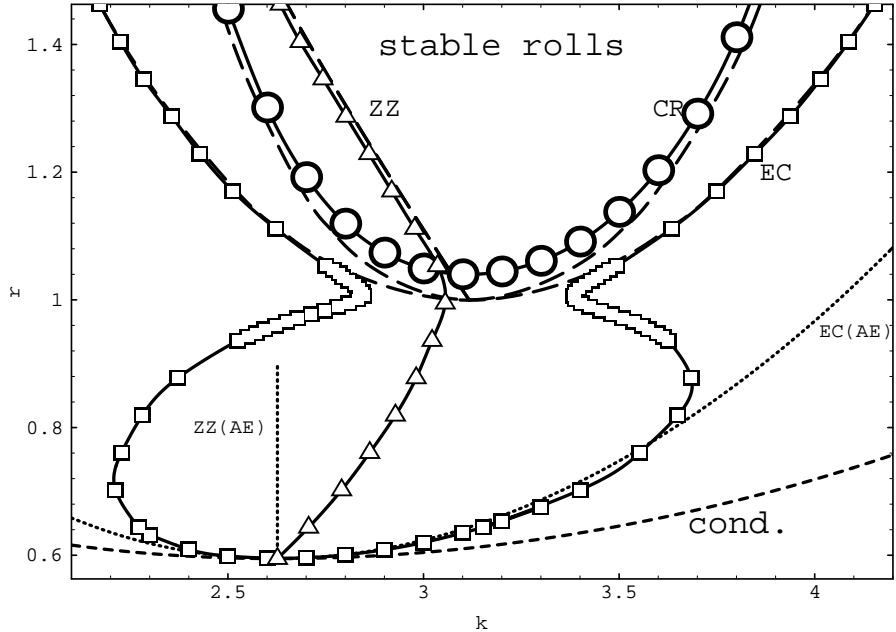}
\end{center}
\caption[]
{Stability boundaries of rolls. Solid lines with open symbols refer to a 
\index{rolls}
\index{stability}
mixture with $Pr=7$, $\psi=0.01$, and $Le=0.025$. The corresponding 
boundaries in a pure fluid with $Pr=7$ are the long dashed curves. Dotted 
curves labelled EC(AE) and ZZ(AE) are predictions of cubic amplitude 
equations for the EC and ZZ boundaries of the mixture, respectively. The
conductive state is stable below the short--dashed curve labelled cond.}
\label{exampleballoon}
\end{figure}
We have calculated all stability boundaries at small $r$ for different values 
\index{stability}
of $Le$, $\psi$, and $Pr$. Concerning the stability behavior of the roll 
\index{rolls}
structures one sees that in the Rayleigh region, $r > 1$, where the 
concentration field is nearly uniform the 
binary mixture behaves like a pure fluid. As for the fixed points, the
transition between Soret
and Rayleigh region is very sharp at small $Le$ and $\psi$. An example for such 
a behavior is given in Fig.~\ref{exampleballoon}. Here only the EC, CR,
and ZZ boundaries are of importance. In the Rayleigh region
of Fig.~\ref{exampleballoon} the CR, ZZ, and EC boundaries of the mixture
(full lines with circles, triangles, and squares, respectively) are lying close
\index{squares}
to the corresponding boundaries of the pure fluid (long-dashed lines). Note 
in particular the vase-like form of the EC boundary $r_{\mathrm{EC}}(k)$ and the 
dent in the ZZ boundary $r_{\mathrm{ZZ}}(k)$: 
Closer to onset ($r_\mathrm{c} \simeq 0.6, k_\mathrm{c} \simeq 2.6$ in 
Fig.~\ref{exampleballoon}), i.e., in the Soret regime $r_{\mathrm{EC}}(k)$ 
opens up 
parabolically and $r_{\mathrm{ZZ}}(k)$ comes out of the critical point linearly
with {\em positive} slope. However, in the crossover range $r \sim 1$ between
Soret and Rayleigh regime the curve $r_{\mathrm{EC}}(k)$ pinches inwards and 
develops a waist such as to follow in the 
Rayleigh regime the parabolic shape of the EC curve of the pure fluid that 
starts out at $k_\mathrm{c}^0 \simeq 3.1, r_\mathrm{c}^0=1$. Similarly 
$r_{\mathrm{ZZ}}(k)$ bends 
in the crossover range towards small $k$ to follow then the ZZ boundary of the 
pure fluid that shows {\em negative} slope.

\begin{figure}[t]
\begin{center}
\includegraphics[width=.9\textwidth]{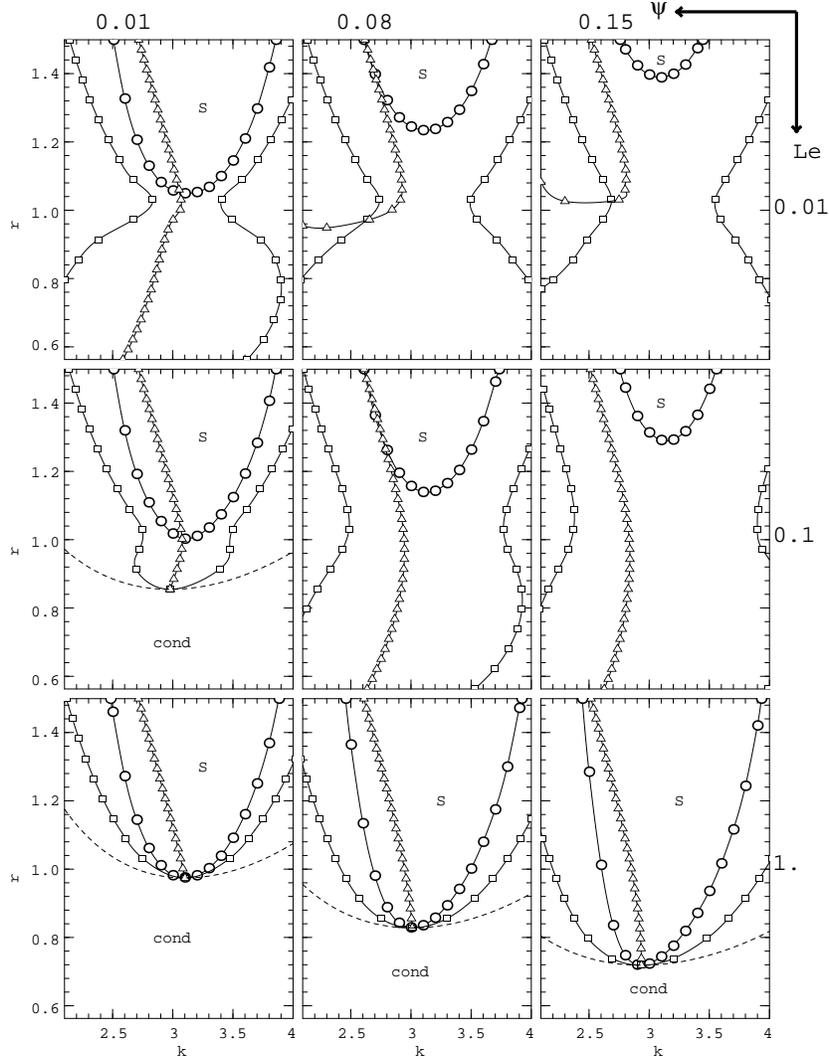}
\end{center}
\caption[]
{Crossections of the stability balloon of rolls in the $k-r$ plane 
\index{rolls}
\index{stability}
at $Pr=10$. Open circles: CR, open squares: EC, open triangles: ZZ. S 
\index{squares}
denotes the region of stable rolls}
\label{balloon10}
\end{figure}

Figure~\ref{balloon10} show that this transition between 
Soret and Rayleigh region that causes the vase-like structure of the EC 
boundary and the sharp bend of the ZZ boundary is smoother for larger $Le$ or
$\psi$. For $Le = 1$, the stability balloon does not qualitatively differ
\index{stability}
from its pure fluid counterpart.

\section{Squares}
\label{squares}
\index{squares}

Convection in square patterns is a fixed point solution that bifurcates out of 
\index{convection}
\index{squares}
the conductive state and that coexists with the roll solution. For parameter 
\index{rolls}
combinations where rolls are not stable at onset, square structures gain
\index{squares}
stability. They lose stability at higher $r$, roughly at the boundary between
\index{stability}
Soret and Rayleigh region. 

\subsection{Technical Remarks}

Squares have the same symmetry plane at $x=0$ as rolls and an 
\index{rolls}
\index{squares}
additional mirror plane at $y=0$. Furthermore, squares
show also a mirror glide symmetry. Here, however, the symmetry transformation
consists of a reflection at
the plane $z=0$ combined with a translation by half a wavelength in 
$x$- {\em and} $y$-direction. To describe these three-dimensional patterns 
the $\Psi$-field
cannot be neglected. We also mention that in contrast to the other fields
 $\Psi$ is odd in $x$ and $y$ and has 
positive parity under the mirror glide operation thereby reflecting the 
symmetries of the velocity field. 
A further reduction of the number of mode occurs since square patterns 
\index{squares}
are invariant under rotation by $90^{\circ}$ mapping the $x$- onto the 
$y$-directions and vice versa. Amplitudes like 
$\Phi_{lmn}$ and $\Phi_{mln}$ are the same. This is also true for $\theta$
and $\zeta$. Again $\Psi$ is different. Here $\Psi_{lmn}=-\Psi_{mln}$.

Because the amount of computational power needed to make a full stability 
\index{stability}
analysis of these three-dimensional structures is too large, we will discuss 
perturbations only for periodic boundary conditions, i.~e.,\ 
$d=b=0$. Here again a separation of perturbations is possible into those 
that change sign or not under the mirror glide operation.
Furthermore, the stability problem is invariant under $x \rightarrow -x$ and
\index{stability} 
$y \rightarrow -y$. Thus one can distinguish between perturbations that are
even in $x$ and $y$, odd in $x$ and $y$, or even in $x$ and odd in $y$ (or
equivalently odd in $x$ and even in $y$). If the perturbations have the same
symmetry in both directions one can in the case of squares finally make use of 
\index{squares}
a last symmetry property and separate between perturbations that are symmetric
or antisymmetric under the exchange of the $x$- and $y$-direction.

The modes taken into account for the Galerkin procedure were chosen in a way 
\index{Galerkin method}
analogous
to the roll case: $X_{lmn}$ and $\delta X_{lmn}$ were neglected, if 
\index{rolls}
$|l|+|m| + n$ was larger than $N_1$ or $N_2$ respectively. Since square 
\index{squares}
structures are stable mainly below 
\index{stability} 
$r=1$ and the boundary layers are less narrow compared to the coexisting roll
solutions (see below), values of $N_2 \leq 20$ were sufficient. 
\index{rolls}

\subsection{Numerical Results}

\subsection*{Square Solutions}
\index{squares}

The convection patterns of squares resemble linear superposition of two 
perpendicular sets of rolls.
\index{rolls}
\index{convection}
\index{squares}
The intensities of the leading modes $\Phi^\mathrm{S}_{101} = 
\Phi^\mathrm{S}_{011}$ of the square solutions show qualitatively the same 
\index{squares}
$r$-dependence as $\Phi^\mathrm{R}_{101}$ for roll solutions but are always 
\index{rolls}
smaller. Arguments based on amplitude equations show however, that 
$|\Phi^\mathrm{S}_{101}|^2 + |\Phi^\mathrm{S}_{011}|^2 > 
|\Phi^\mathrm{R}_{101}|^2$ near onset if and only if squares are stable there. 
\index{squares}
\index{stability}
But this is no longer true at higher $r$ where the solutions are such that
$|\Phi^\mathrm{S}_{101}|^2 + |\Phi^\mathrm{S}_{011}|^2 < 
|\Phi^\mathrm{R}_{101}|^2$. This change takes place before squares lose 
\index{squares}
stability. 

\begin{figure}[t]
\begin{center}
\includegraphics[width=.9\textwidth]{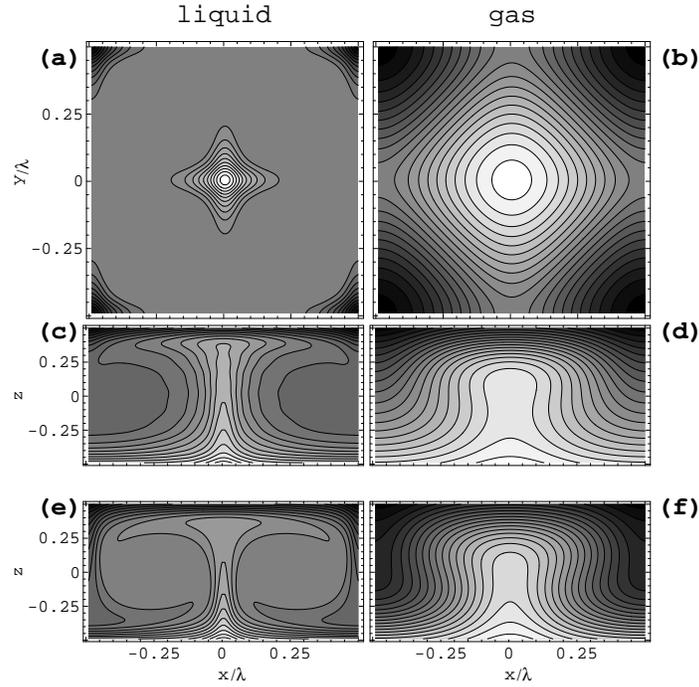}
\end{center}
\caption[]
{Structural properties of square (a-d) and roll (e,f) convection
\index{squares}
\index{rolls}
\index{convection}
for representative liquid parameters ($Le = 0.01$, $Pr = 10$, left column)
and gas parameters ($Le = Pr = 1$, right column) at $r=1, \psi=0.15$.
In (a,b) the concentration distribution of squares at mid height, $z=0$, 
\index{squares}
is shown. In (c-f) we show the concentration distribution in
a vertical cross section at $y=0$. The largest vertical upflow is at
$x=y=0$} 
\label{shadowgraph}
\end{figure}
In Fig.~\ref{shadowgraph} we show the concentration distribution 
of square convection for two parameter combinations that are representative
\index{squares}
\index{convection}
for liquid and gas mixtures. This plot and the concentration 
field structure of 
rolls and squares in a vertical cross section shows a characteristic boundary
\index{squares}
\index{rolls}
layer and plume behavior at small $Le$. Such structures occur when 
advective mixing is large compared to diffusion in the bulk of the fluid.
Consequently the boundary layers and plumes are more pronounced in rolls than 
\index{rolls}
in squares since the leading velocity amplitudes are greater for the former.
\index{squares}
Thus squares with their broader boundary layers are much smoother structures 
than rolls at the same parameters. The practically harmonic velocity and 
\index{rolls}
temperature fields are not shown. 

\subsection*{Stability Properties of Squares}
\index{squares}

Performing the stability analysis of squares we had to restrict ourselves to the
case $d=b=0$ where the perturbations separate into different symmetry
classes. 
Because both squares and rolls can be described as even in $x$ and
\index{rolls}
\index{squares}
$y$ and as mirror glide antisymmetric, one expects a perturbation that
destabilizes the squares and favors the rolls to fulfill these symmetries, too.
\index{stability}
\index{squares}
However such a perturbation should break the symmetry $x \leftrightarrow y$. 
We actually always found the most critical perturbation to fall into this
symmetry class. Other perturbations that break the mirror symmetry in 
$x$- or $y$-direction are less critical. 

\begin{figure}[t]
\begin{center}
\includegraphics[width=.9\textwidth]{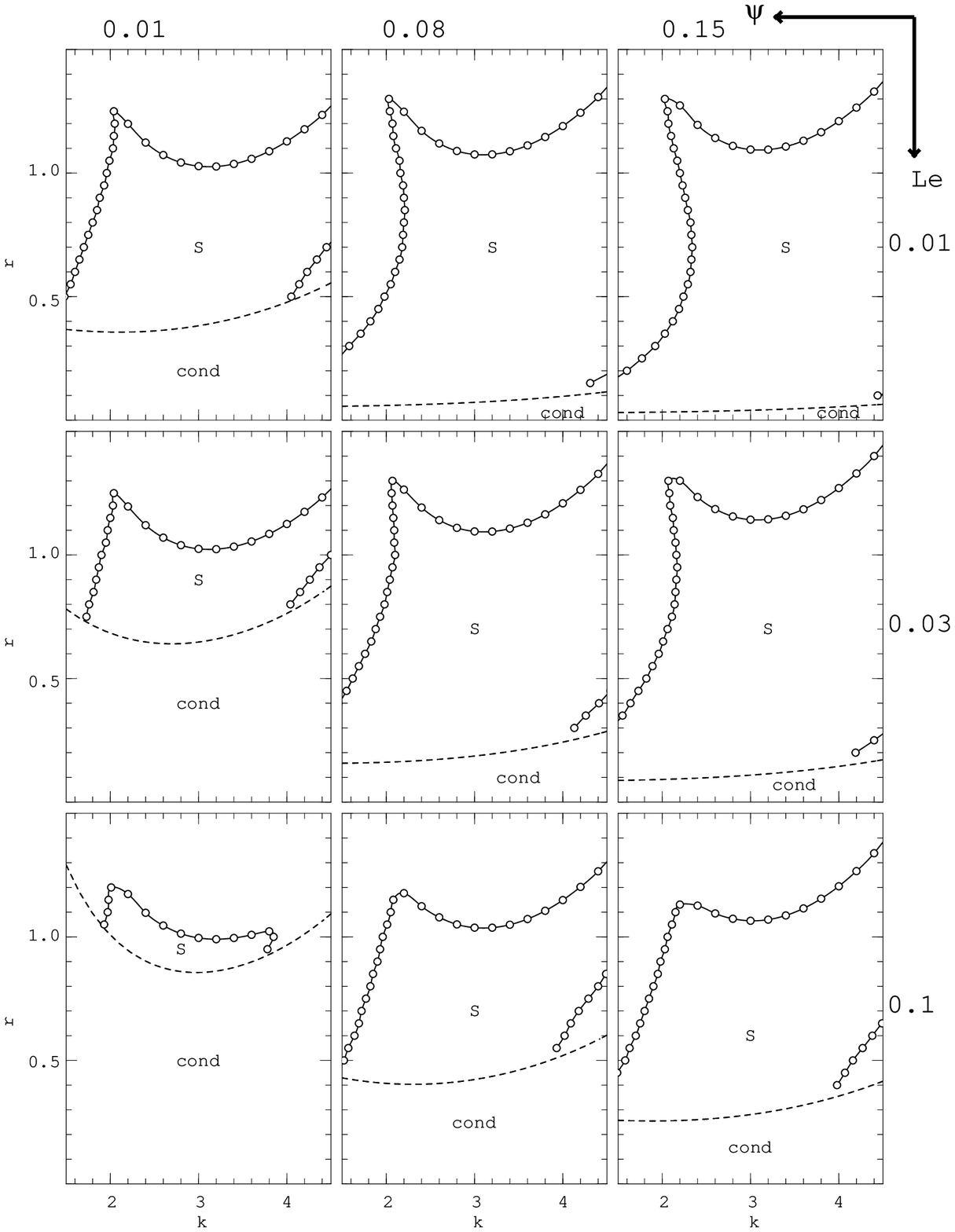}
\end{center}
\caption[]
{Crossections of the stability balloon of squares in the $k-r$ plane for 
\index{squares}
\index{stability}
$Pr=10$ obtained from a restricted stability analysis as explained in the 
text. S denotes the region of stable squares}
 \index{squares}
\label{squareballoon}
\end{figure}
Figure~\ref{squareballoon} shows typical examples for the stability
\index{stability}
region of squares. The left and right side of the stability boundaries should 
\index{squares}
not be taken too serious -- presumably square structures are destabilized
earlier by instabilities with finite $b$ or $d$ that tune the wavenumber and 
\index{stability} \index{instability}
that are not considered here. 

Even if squares are stable at onset they always lose  their stability
\index{stability}
\index{squares}
against a roll pattern at higher $r$. Furthermore, for certain parameters 
\index{rolls}
\index{squares}
there does also exist a band of $r$-values where neither squares nor rolls 
are stable. Therein crossrolls or oscillations appear as stable patterns.
\index{stability}
\index{crossrolls}  
   
\section{Crossrolls}
\label{crossrolls}
\index{crossrolls} 

Crossrolls are 3D convection patterns that show the same symmetries as
\index{convection}
\index{crossrolls} 
squares except the $x \leftrightarrow y$ symmetry. As squares, they can be
\index{squares}
qualitatively described as superpositions of perpendicular roll patterns.
\index{rolls}
But in the case of crossrolls the amplitudes $\Phi_{101}$ and $\Phi_{011}$
of the $x$- and $y$-rolls are not the same.
\index{crossrolls}

Crossroll structures appear in two modifications: as stationary and as
oscillatory crossrolls. For the latter, the amplitudes of $x$- and 
$y$-rolls are time dependent.
\index{rolls}
\index{crossrolls}

\subsection{Technical Remarks}

The numerical investigations of the crossroll structures were performed as in 
\index{crossrolls}
the case of squares except that modes like $X_{lmn}$ and $X_{mln}$ could not be
\index{squares}
identified anymore due to the lack of the $x \leftrightarrow y$ symmetry.
For $b=d=0$ instabilities are either mirror glide symmetric or antisymmetric.
\index{stability} \index{instability}
They are either even or odd in the $x$- and the $y$-direction. We used
expansions up to $N_2 = 20$ to investigate these structures.
The time-dependent structures were investigated via direct numerical
integration of the equations of motion for the modes. 

\subsection{Numerical Results}

\begin{figure}[t]
\begin{center}
\includegraphics[width=.9\textwidth,angle=270]{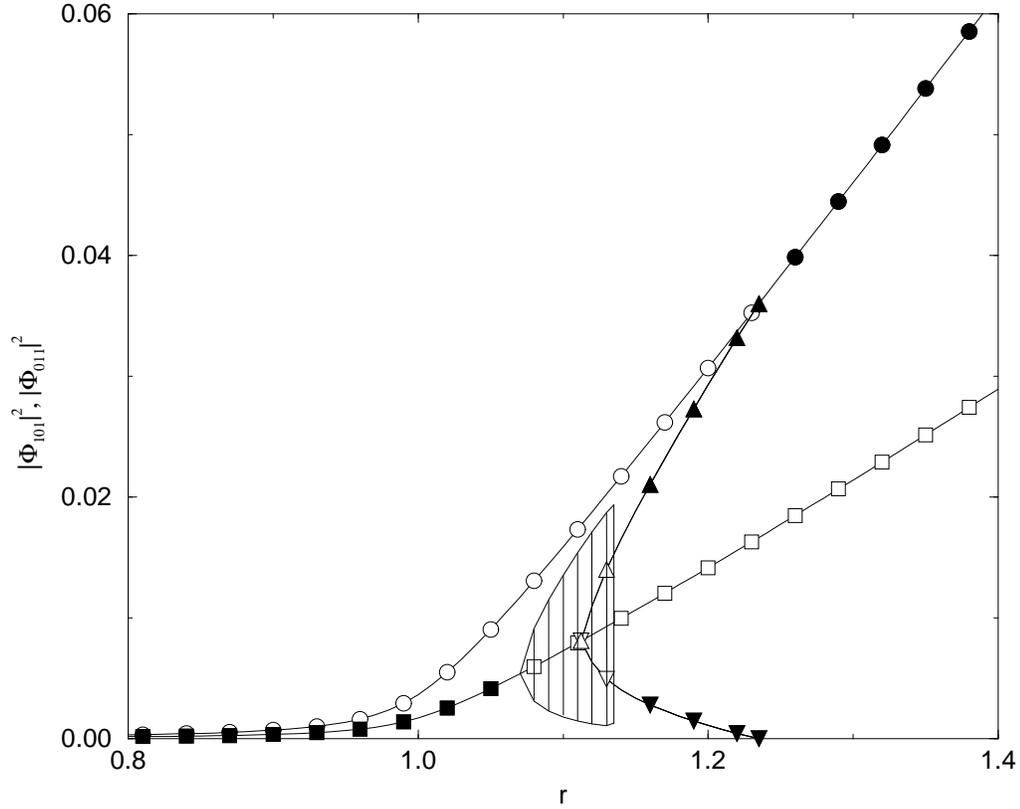}
\end{center}
\caption[]
{Bifurcation diagram of $x$-roll intensity and $y$-roll intensity
\index{rolls}
versus $r$ for squares (squares), rolls (circles), stationary  
\index{rolls}
\index{squares}
(triangles), and oscillatory crossrolls. Filled (open) 
\index{crossrolls}
symbols denote  
stable (unstable) states. Full lines delimiting the hatched area 
\index{stability}
mark maxima and minima of oscillations of $|\Phi_{101}|^2$ and 
$|\Phi_{011}|^2$. 
The parameters are \cite{MS91} $\psi= 0.23, L = 0.0045, Pr = 27$, and
$k=\pi$} 
\label{bif}    
\end{figure}

Stationary crossrolls exist at parameters 
\index{crossrolls}
$Pr,L,\psi$ if squares are stable at onset. Then, they appear in the 
\index{squares}
$r$-range were squares are already and rolls are still unstable. Oscillatory
\index{rolls}
\index{stability}
crossrolls exist only at small $Le$. The connection between the different
\index{crossrolls}
structures can be extracted from the bifurcation diagram in Fig.~\ref{bif}.
For the parameters $Pr=27$, $Le=0.0047$, and $\psi = 0.23$ taken from
\cite{MS91}, all four kinds of patterns exist. For the calculation of the fixed
points a model with $N_2 = 10$ was used, the bifurcation diagram is therefore
only qualitatively correct at higher $r$.

\subsection*{Stationary Crossrolls}
\index{crossrolls} 

We will first ignore the crossroll solutions of the oscillatory type and
concentrate on the stationary patterns.
At onset squares with $|\Phi_{101}|^2 = |\Phi_{011}|^2$ are the stable 
\index{squares}
structure. The coexisting roll branch is unstable. At about $r=1.11$ a new
\index{rolls}
\index{stability}
type of solutions bifurcates out of the square branch, the stationary 
crossrolls. In this $r$-range neither squares nor rolls are stable. 
\index{crossrolls} 
\index{rolls}
\index{squares}
For the crossroll structures the two important mode intensities are different,
\index{crossrolls}
say $|\Phi_{101}|^2 > |\Phi_{011}|^2$. The difference between these intensities
grows until $|\Phi_{011}|^2 = 0$ and $|\Phi_{101}|^2$ touches the roll branch
\index{rolls}
at another bifurcation point at $r \approx 1.24$. The crossrolls 
transfers their stability to the rolls here. 
\index{rolls}
\index{stability}
\index{crossrolls}

The stationary crossrolls exist always inside an intermediate $r$-range if 
squares are stable at onset. The oscillatory type appears only at very small 
\index{squares}
\index{stability}
\index{crossrolls}
$Le$ as in Fig.~\ref{bif}. If these solutions are absent, the squares 
transfer their stability directly to the stationary crossroll branch, leading 
\index{stability}
\index{crossrolls}
to a sequence squares -- stationary crossrolls -- rolls of stable structures
\index{rolls}
\index{squares}
for increasing $r$. In particular, the stability boundaries that limit the
region of stable squares in Fig.~\ref{squareballoon} are all of the type
leading to stationary rolls: oscillatory crossrolls do not exist for these
parameters.
\index{squares}
\index{stability}
\index{crossrolls}

\subsection*{Oscillatory Crossrolls}
\index{crossrolls}

But for the parameters in Fig.~\ref{bif} the bifurcation behavior is more
complicated. The squares lose their stability already at $r \approx 1.08$ 
\index{stability}
\index{squares}
before the stationary crossrolls emerge. At this point oscillatory crossrolls
grow out of the square state in a supercritical Hopf bifurcation. The 
\index{squares}
stationary crossrolls that appear at $r=1.11$ remain unstable until
$r \approx 1.14$.
\index{crossrolls}

\begin{figure}[t]
\begin{center}
\includegraphics[width=.9\textwidth]{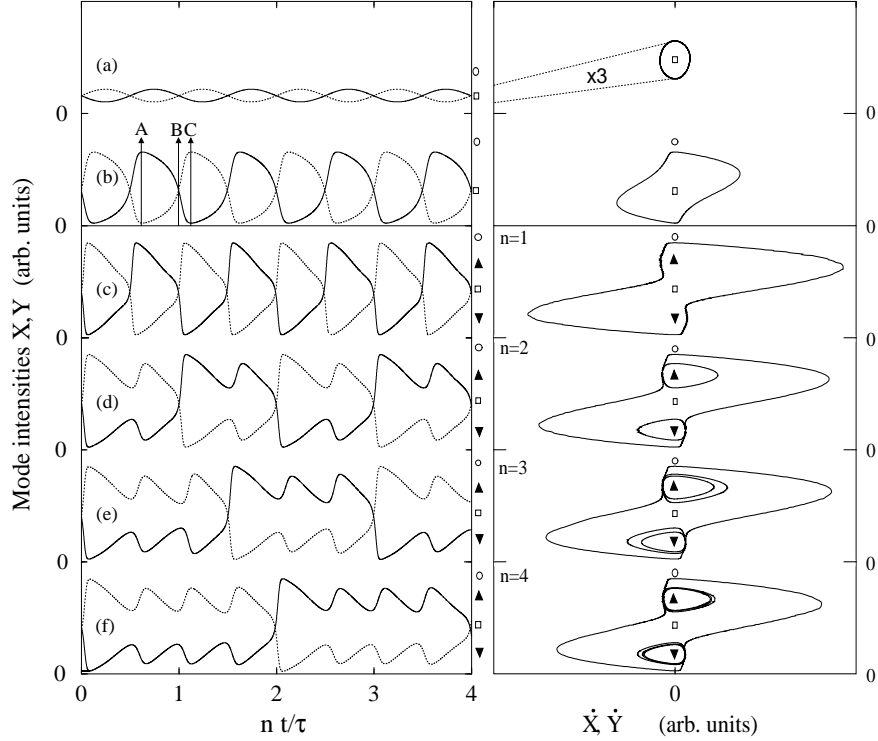}
\end{center}
\caption[]
{Evolution of the oscillatory dynamics with increasing $r$  
from top to bottom. Parameters as in Fig.~\ref{bif}. 
(a) and (b) are states located shortly above the Hopf threshold and right 
below the crossroll bifurcation, respectively. (c) -- (f) are SC states. 
\index{crossrolls}
The right column contains phase space plots of $X \sim |\Phi_{101}|^2$ and 
$Y \sim |\Phi_{011}|^2$ versus their time derivatives for the intensities
$X$, $Y$ shown in the left column with arbitrary 
common units versus reduced time $n t/\tau$. In the SC states $n$ 
is the number of windings around each of the stable crossroll fixed points 
(triangles). Squares and circles indicate unstable square and roll states,
\index{rolls}
\index{squares}
\index{stability}
respectively} 
\label{os2}    
\end{figure}

Fig.~\ref{os2} shows  for several Rayleigh numbers 
that $|\Phi_{101}|^2(t)$ and $|\Phi_{011}|^2(t)$ oscillate in opposite phase 
around a common mean value given by the unstable square state. 
\index{squares}
\index{stability}
The $x$-roll intensity of the pulsating pattern grows or decreases on 
cost of the $y$-roll intensity, however, without ever going to zero. Thus 
\index{rolls}
the two roll sets never die out completely or reverse their turning  
direction during the oscillations. 
Close to the Hopf bifurcation the oscillations are harmonic and of  
small amplitude (Fig.~\ref{os2}a). With increasing $r$ the  
frequency decreases roughly linearly \cite{MS91}. Furthermore, and more 
importantly, with increasing amplitude the oscillation becomes more 
anharmonic and relaxational in Fig.~\ref{os2}b, c; see also 
Ref.~\cite{GPC85,MS91}: While the system rapidly sweeps through the square 
\index{squares}
state it spends more and more time in the vicinity of the roll state. 
\index{rolls}
The change from harmonic in Fig.~\ref{os2}a to  
strongly relaxational oscillations in Fig.~\ref{os2}c is also 
documented in the right column of Fig.~\ref{os2}. There we show phase 
space plots associated with the time histories of  
$|\Phi_{101}|^2$, $|\Phi_{011}|^2$ in the left column. 
 
At larger $r$ the system gets attracted into one of the crossroll fixed points 
that have become stable shortly before oscillatory crossrolls cease to exist 
\index{crossrolls}
\index{stability}
in Fig.~\ref{bif}. In this interval we 
have observed a novel subharmonic bifurcation cascade (SC) in which 
the crossroll attractors entrain the oscillations: First the crossroll 
\index{crossrolls}
attractors deform the phase trajectory (Fig.~\ref{os2}c). Then with 
increasing $r$ the oscillations execute an increasing number of windings  
around the crossroll states. In Figs.~\ref{os2}c -- f 
the winding number around a crossroll fixed point increases from $n$ = 1 to 
$n$ = 4 and the  period $\tau$ of the oscillations increases from  
$2 \pi/\omega$ to $2\pi n/\omega$ in integer 
steps. This increase of the winding number in the SC continues 
beyond $n=4$; we  have found also $n = 5$. However, the control parameter 
interval  $\delta r_n$ for an $n$-cycle becomes so narrow --
$\delta r_n \cong (1.6, 0.6, 0.2) 10^{-4}$ 
for $n=(2, 3, 4)$ -- that our numerical resources were not  
sufficient to resolve the SC further. But we think that the SC is a robust, 
experimentally accessible phenomenon;  
Ref.~\cite{GPC85} contains a hint for a 2-cycle. Beyond the 
SC interval in the system gets   
attracted into one of the crossroll fixed points. The transition  between the 
oscillations  and the crossroll state is slightly hysteretic. Upon reducing 
\index{crossrolls}
$r$ the system remains in the crossroll state until below the SC interval and 
then a transition to an $n = 1$ oscillation occurs. So in a small 
$r$-interval there is bistable coexistence of oscillations and  
stationary crossrolls. 
\index{stability}
\index{crossrolls}

\section{Conclusion}
\label{conclude}

We reviewed our findings of Rayleigh--B\'enard convection in
\index{Rayleigh--B\'enard system}
\index{convection}
binary mixtures with positive Soret effect. In this case the lighter 
component of the mixture is driven into the direction of higher temperature. 
The enhanced density gradient leads to a convective instability at smaller 
\index{stability} \index{instability}
temperature differences than in the pure fluid case.
Convection at positive $\psi$ shows a rich variety of stable structures at
moderate $r \leq 1.5$. 

Using the Galerkin technique, we investigated two- and
\index{Galerkin method}
three-dimensional laterally periodic patterns as rolls, squares, and
\index{rolls}
\index{squares}
stationary and oscillatory crossrolls and determined their stability behavior. 
\index{stability}
\index{crossrolls}
All these patterns are realized as stable convection structures somewhere in 
parameter space. 
The numerical examination of convection in liquid binary mixtures is in general 
\index{convection}
much more difficult compared to pure fluids since small values of the Lewis
number $Le \leq 0.01$ lead to a pronounced narrow boundary layer behavior of
the concentration field. This is especially the case for rolls whereas squares
\index{rolls}
\index{squares}
are somewhat smoother. Nevertheless we had to restrict the stability analysis 
\index{stability}
of the three-dimensional structures to perturbations with the same wave 
number.

Squares and rolls coexist as convective solutions above onset. The square 
\index{rolls}
\index{squares}
symmetry requires that the main velocity modes are equal: 
$|\Phi_{101}| = |\Phi_{011}|$. For the two-dimensional rolls on the other hand
\index{rolls}
one of these amplitudes, say $|\Phi_{011}|$ is zero. Squares are stable in the 
Soret region if $Le$ is sufficiently small. They transfer their stability to 
crossrolls at higher $r$. If $Le$ is not too small, only stationary crossrolls
\index{crossrolls} 
exist. As squares, they can be described as linear superpositions of roll
\index{rolls}
\index{squares}
structures. For them however, $|\Phi_{101}| > |\Phi_{011}|$ or vice 
versa. With increasing $r$ the smaller amplitude tends to zero until the
crossroll branch ends on the now stable roll branch in the bifurcation diagram. 
\index{rolls}
\index{stability}
\index{crossrolls}

At very small $Le$ there exists also a branch of oscillatory crossrolls that
\index{crossrolls} 
emerges out of the square branch in a supercritical Hopf bifurcation at smaller 
\index{squares}
$r$ than the branch of the stationary type. In the oscillatory crossrolls
\index{crossrolls}     
$|\Phi_{101}|^2(t)$ and $|\Phi_{011}|^2(t)$ oscillate in opposite phase 
around a common mean value given by the unstable square state. At higher $r$ the
oscillatory crossrolls disappear in a subharmonic bifurcation cascade and the
stationary ones gain stability.
\index{stability}
\index{crossrolls}     

At higher $Le$ roll structures are stable directly above onset and crossrolls
\index{rolls}
structures do not appear.
\index{crossrolls}     
The analysis of the rolls shows that in the explored parameter range
\index{rolls}
only the basic mechanisms of instability 
\index{stability} \index{instability}
occur that are already known from the pure fluid, namely the Eckhaus, zigzag, 
crossroll, oscillatory, and skewed varicose mechanism. When the Soret region is 
small, the stability balloon of the mixtures resembles the Busse balloon for
\index{stability}
the pure fluid. However, at small $Le$
when the Soret region is large the situation is different. The fixed point 
solutions show a sharp transition between the two regimes. The convection 
\index{convection}
amplitudes are very small in the Soret regime. But near $r =1$ they increase 
strongly and become comparable to those of the pure 
fluid convection. The stability boundaries of roll convection show a similar 
\index{convection}
\index{rolls}
\index{stability}
transition here. In the Rayleigh region the boundaries are close to the 
boundaries of the pure fluid. But upon reducing the Rayleigh number
the boundaries begin to deviate from their pure fluid 
counterparts. Near onset they finally agree with the predictions of the 
amplitude equations for the mixtures.

\section*{Glossary}
\begin{tabular}{ll}
$b$ & wave vector deviation in $y$\\
$c$ & concentration deviation\\
$C$ & concentration\\
$C_0$ & mean concentration \\
$C_\mathrm{cond}$ & concentration in conductive state \\
CR & crossroll instability\\
$d$ & wave vector deviation in $x$, also plate distance \\
${\bf e}_x, {\bf e}_y, {\bf e}_z$ & unit vectors \\
EC & Eckhaus instability\\
$\vec{g}$ & gravitational field \\
$k$ & wave number \\
$k_x$ & wave number in x \\
$k_y$ & wave number in y \\
$k_\mathrm{c}$ & critical wave number \\
$k_\mathrm{c}^0$ &critical wave number for pure fluid\\
OS & oscillatory instability \\
$p$ & pressure deviation\\
$P$ & pressure \\
$r$ & reduced Rayleigh number \\
$Ra_\mathrm{c}$ & critical Rayleigh number \\
$Ra_\mathrm{c}^0$ & critical Rayleigh number for pure fluid\\
$s$ & growth rate \\
SV & skeved varicose instability \\
$T_0$ & mean temperature\\
$T_\mathrm{cond}$ & temperature in conductive state \\
$\Delta T$ & temperature difference\\
$u$ & velocity, x component \\
$v$ & velocity, y component \\
$w$ & velocity, z component \\
$X$ & arbitrary field \\
$X_{lmn}$ & field mode \\
$\delta X$ & perturbation in field $X$ \\
$\delta X_{lmn}$ & perturbation mode \\
ZZ & zigzag instability \\
$\zeta$ & $c - \psi \theta$\\
$\theta$ & temperature deviation \\
$\nu$ & kinematic viscosity\\
$\rho_0$ & mean density\\
$\Phi$ & velocity potential \\
$\tau$ & period of oscillations\\
$\Psi$ & velocity potential \\
$\omega$ & frequency of oscillations \\
\end{tabular}
\end{document}